\documentclass[a4paper]{jpconf}

\newlength{\vshift}
\newlength{\hshift}
\usepackage{amssymb,amsopn}
\usepackage{slashed}
\usepackage[utf8]{inputenc}
\usepackage{amsmath}
\usepackage{graphicx}
\usepackage{color}

\def\d{\textrm{d}}
\def\ds{\stackrel{\star}{,}}

\def\nn{\nonumber}
\def\be{\begin{equation}}             \def\ee{\end{equation}}
\def\ba#1{\begin{array}{#1}}          \def\ea{\end{array}}
\def\bea{\begin{eqnarray} }           \def\eea{\end{eqnarray} }
\def\beann{\begin{eqnarray*} }        \def\eeann{\end{eqnarray*} }
\def\beal{\begin{eqalign}}            \def\eeal{\end{eqalign}}
             
\def\bsubeq{\begin{subequations}}     \def\esubeq{\end{subequations}}
\def\bitem{\begin{itemize}}           \def\eitem{\end{itemize}}

\def\m{\mu}

\begin{document}
\title{Search for footprints of quantum spacetime in black hole QNM spectrum}

\author{Marija Dimitrijevi\'{c} \'{C}iri\'{c} $^1$, Nikola Konjik$^2$ 
 and Andjelo Samsarov$^{3}$}
\address{$^{1,2}$ Faculty of Physics, University of Belgrade, Studentski trg 12,
11000 Beograd, Serbia }
\address{$^3$ Rudjer Bo\v skovi\'c Institute, Theoretical Physics Division, Bijeni\v cka  c.54, HR-10002 Zagreb, Croatia}
\ead{$^1$ dmarija@ipb.ac.rs, $^2$  konjik@ipb.ac.rs, $^3$  asamsarov@irb.hr}                        


\begin{abstract}
Black hole (BH) perturbation  is followed by a ringdown phase which is dominated by quasinormal modes (QNM). These modes may provide key signature of the gravitational waves. The presence of a deformed  spacetime   structure  may  distort this signal. In order to account for such  effects, we consider  a toy model consisting of a noncommutative charged scalar field propagating in a  realistic black hole background. We then analyse the corresponding field dynamics by applying the methods of the Hopf algebra deformation by Drinfeld twist.   The latter framework is well  suited for incorporating  deformed symmetries  into a study of this kind. As a result, we obtain the BH QNM spectrum that, besides containing the intrinsic information  about  a black hole that is being analysed,  also carry the information about the underlying   structure of spacetime.
\end{abstract}

\section{Introduction}
     
Many different approaches to quantum gravity suggest that  a continuous differential manifold  does not provide  an adequate description of spacetime at very small length scales.
Indeed, probing ever smaller distances of space requires more and more energetic particles which sooner or later reach their energy threshold that sets a natural bound on the process of spacetime inquiry. Whilst this energy threshold is being given by the Planck mass, its very exitence  points toward a quantum nature of spacetime.

At the same time a natural question  arises as to how could one experimentally verify that the actual structure of spacetime is quantized. One route  to verify this extremely challenging premise involves quantum spacetime and quantum gravity phenomenology
where there has been a long standing interest in the possibility of Planck scale departures from Lorentz symmetry \cite{AmelinoCamelia:2009pg},\cite{AmelinoCamelia:2008qg},\cite{jacobson}. In particular, 
a research in quantum gravity phenomenology has focused on the question of the fate
of Lorentz invariance largely through a perspective of modified energy-momentum dispersion relations. Over the  years several scenarios have arisen for dispersion of light motivated
by theories and hypotheses about quantum gravity \cite{amelinoellismavro},\cite{gambinipullin}.

These include a study of many different aspects among which we  mention various scenarios of variable speed of light (VSL) \cite{schaefer},\cite{biller},  corrections to neutrino propagation \cite{urrutia},\cite{Amelino-Camelia:2016fuh}, different schemes for implementing a relativity theory with minimum length \cite{dsramelino},\cite{dsramelino1},\cite{kowalski},\cite{bruno},\cite{magueijo}, the principle of relative locality \cite{AmelinoCamelia:2011bm} and time delay in arrival of photons with different energy \cite{AmelinoCamelia:2011cv},\cite{Meljanac:2012pv},\cite{Mignemi:2016ilu},\cite{Mignemi:2018tdg}. There are diverse theoretical frameworks  \cite{KowalskiGlikman:2002we},\cite{KowalskiGlikman:2002jr},\cite{Borowiec:2008uj},\cite{Borowiec:2010yw},\cite{Juric:2015jxa},\cite{Aschieri:2017ost} based on noncommutative (NC) spacetime which give rise to modified dispersion relations.

Although all physical effects predicted along the lines just described are really tiny,
 one can still hope to get a meassurable outcome. The crucial point in this and in all such kind of experiments in general,  is that the effects are cumulative. Thus for example, in the case of time delay in arrival of two light beams of different frequencies emitted by gamma ray bursts,
the photons  created in the process  travel across the space    and while    traversing  large cosmic distances, they  accumulate
all these tiny little outcomes in a manner similar to a snowball rolling down a snowy hill,
 so that by the time they reach detector, the total effect of time delay becomes observable.

The idea of probing the physics of quantum gravity with high energy astrophysical
observations has been out there for quite some time already, but with the launch of the
Fermi gamma ray telescope \cite{abdo} it stepped out into a new era with a  lot more possibilities.

Another intriguing possibility for trapping any potential signal  of a quantized structure of spacetime may eventually  be found in the quasinormal mode (QNM) spectrum of a generic black hole exhibited during its relaxation phase, when it goes through a so called ringdown period of quasinormal ringing. During this period  black holes emit gravitational waves, 
 which became accessible to an experimental inquiry since their first detection ever took place in the recent LIGO experiment \cite{LIGO}.
In essence the purpose of the papers \cite{Ciric:2017rnf},\cite{Ciric:2019uab}, as well as of the present paper is to investigate this idea on the example of a realistic $4$-dimensional  black hole \footnote{The effect of noncommutativity on the QNM spectrum of $3$-dimensional asymptotically adS spacetimes were investigated in \cite{Gupta:2015uga},\cite{Gupta:2017lwk}.}. Specifically, besides assuming that  quantum nature of spacetime may  leave its trace in the QNM spectrum  of a black hole, we also propose a specific model which enable us to sort out
the trace of quantum spacetime
 and quantify analytically the associated  signatures of spacetime noncommutativity.  

Quasinormal modes \cite{rg},\cite{vish},\cite{press},\cite{chandra},\cite{mashoon},\cite{nollert-review},\cite{Kokkotas:1999bd},\cite{cardosoreview},\cite{KonopReview},\cite{Cardoso:2001hn},\cite{Cardoso:2001bb} dominate the relaxation phase of the black hole perturbation dynamics,
which according to the uniqueness and no-hair theorems \cite{israel},\cite{carter},\cite{hawking},\cite{robinson},\cite{isper},  after some time acquires a form of the Kerr-Newman geometry characterized solely by the black-hole mass, charge, and angular momentum. Moreover, quasinormal modes accommodate features enclosed by these theorems, in particular the property
that in asymptotically flat spacetimes, spherically symmetric charged black holes cannot support any external static matter configurations
made of charged massive scalar fields \cite{bekenstein}. As this implies that
perturbation fields left outside the newly created black hole would either be radiated away to infinity or swallowed by the
black hole, the boundary conditions defining QNMs comply with the requirement of behaving as
 purely outgoing waves at spatial infinity and purely ingoing waves crossing the event horizon.


With a purpose of studying the signature of quantum spacetime as revealed within a QNM spectrum, we consider the nearly spherical gravitational collapse of a charged scalar field to form a charged Reissner-Nordstr\"om black hole. This we achieve by studying  dynamics of the charged massive scalar perturbation field. The focus would be on the relaxation phase of the charged perturbation fields which were left outside the newly created black hole.

Quantum structure of spacetime may be implemented in various ways, with noncommutative deformation being one of the most notable ones. The noncommutativity itself can be as well  introduced  in many different ways \cite{connes},\cite{landi},\cite{madore}. Here we follow the twist approach \cite{NCbookMi} and deform the Poincar\'e algebra to a twisted Poincar\'e algebra, so that the whole deformation is squeezed into a coalgebraic sector, while the algebra remains unchanged.
 The change  in the coalgebra is relavant for multiparticle states \cite{Daszkiewicz:2007az},\cite{Daszkiewicz:2007ru},\cite{Young:2007ag},\cite{Young:2008zm},\cite{Govindarajan:2009wt},\cite{Meljanac:2011mt},\cite{Meljanac:2012fa},\cite{Ivetic:2016qtz},\cite{Napulj2018}. One of the adventages of the twist deformation is that it induces a deformed differential calculus in a well defined way. In particular, it introduces a deformed product of functions, the $\star$-product, and more generally, deformed tensor product between tensor fields, including  deformed wedge product $\wedge_\star$ between differential forms.

 In order to
realize this approach, we choose a twist operator of the form 
\begin{eqnarray}
\mathcal{F} &=& e^{-\frac{i}{2}\theta ^{\alpha\beta}\partial_\alpha\otimes \partial_\beta} \nn\\
&=& e^{-\frac{ia}{2} (\partial_t\otimes\partial_\varphi - \partial_\varphi\otimes\partial_t)}.
\label{AngTwist0Phi}
\end{eqnarray} 
Vector fields $X_{1}=\partial _t$,
$X_{2}= \partial_\varphi$ are commuting vector fields,
$[X_{1},X_{2}]=0$, therefore the twist (\ref{AngTwist0Phi}) is an Abelian twist \cite{PL09}. It is dubbed as "angular twist" becuase the vector field $X_2 =
\partial_\varphi$ is a generator of rotations around $z$-axis. 
The twist (\ref{AngTwist0Phi}) defines the $\star$-product of functions,
\begin{eqnarray}
f\star g &=&  \mu \{ e^{\frac{ia}{2} (\partial_t\otimes \partial_\varphi - \partial_\varphi\otimes
\partial_t)}
f\otimes g \}\nn\\
&=& fg + \frac{ia}{2}
(\partial_t f(\partial_\varphi g) - \partial_t g(\partial_\varphi f)) + 
\mathcal{O}(a^2) .\label{fStarg0Phi}
\end{eqnarray}


\section{Introducing the model}

   As indicated in the previous section, in order to study a ringdown phase, that is a relaxation dynamics of  charged matter around a newly formed black hole horizon, we consider the action 
\begin{eqnarray}
S[\hat{\phi}, \hat{A}] &=& \int {\Big( \d {\hat{\phi}} - i\hat{A} \star {\hat{\phi}} \Big)}^+
\wedge_\star  *_H \Big( \d \hat{\phi} -i\hat{A} \star \hat{\phi} \Big) \nonumber \\
&& - \int \frac{\mu^2}{4!} \hat{\phi}^+\star \hat{\phi} \epsilon_{abcd}~ e^{a} \wedge_\star
e^b\wedge_\star e^c \wedge_\star e^d
- \frac{1}{4 q^2} \int (*_H \hat{F}) \wedge_\star \hat{F}. \label{NCActionGeometric}
\end{eqnarray}
that governs dynamics of a massive, charged scalar field $\hat{\phi}$  in some gravity background (which, unlike the scalar and gauge field, is supposed to be fixed and not treated as a dynamical degree of freedom, see down below). 
The scalar field  $\hat{\phi}$ has the mass $\mu$ and charge $q$ and  transforms in the fundamental representation of NC $U(1)_\star$. 
 The one-form  noncommutative (NC) $U(1)$  gauge field $\hat{A}=\hat{A}_\mu \star \d x^{\mu}$ is introduced in the model  through a minimal coupling. Likewise,  the two-form
field-strength tensor is defined as
\begin{equation}
\hat{F} = \d \hat{A} - \hat{A} \wedge_\star
\hat{A} =\frac{1}{2} \hat{F}_{\m\nu}\star \d x^{\mu} \wedge_\star \d
x^{\nu}. \label{NCF}
\end{equation}

The presence of  deformed structural maps in the action (\ref{NCActionGeometric}), i.e. deformed wedge product  $\wedge_\star$ and star product (\ref{fStarg0Phi}), together with the fact that  (\ref{NCActionGeometric})  is being written in terms of a maximal volume form,  ensures that
the action  (\ref{NCActionGeometric}) is invariant under twisted diffeomorphisms.   Note however that we work in a fixed geometry, therefore the diffeomorphisms are limited to those that preserve a given fixed background metric.
In a special case of
a flat Minkowski space-time, without any coupling to gravity, a corresponding form of the action (\ref{NCActionGeometric}) is invariant under a  twisted Poincare algebra which is obtained by twisting the original  Poincar\'e algebra with the particular type of Drinfeld twist operator (\ref{AngTwist0Phi}). Of course, the presence of deformed structural maps in the action (\ref{NCActionGeometric})  also ensures the implementation of quantum deformation techniques in the theory.

In order that the mass term for the scalar field $\hat{\phi}$ takes on a geometrical form, we had to
introduce the vierbein one-forms $e^a=e^a_\mu\star \d x^{\mu},$ which satisfy $g_{\mu\nu} = \eta_{ab}e_\mu^a\star
e_\nu^b$. Here we point out that the metric $g_{\mu \nu}$ and vierbeins in the action (\ref{NCActionGeometric})  are not treated as  dynamical variables. Instead, the metric is fixed to be that of the
Reissner-Nordstr\"{o}m (RN) type,
\begin{equation}
{\rm d}s^2 = (1-\frac{2MG}{r}+\frac{Q^2G}{r^2}){\rm d}t^2 - \frac{{\rm
d}r^2}{1-\frac{2MG}{r}+\frac{Q^2G}{r^2}} - r^2({\rm d}\theta^2 + \sin^2\theta{\rm d}\varphi^2), \label{dsRN}
\end{equation}
where $M$ and $Q$ are the mass   and charge of the RN black hole, respectively.
In view of that, the vector fields $X_1$ and $X_2$ are two Killing vectors for this metric. In particular, the twist
(\ref{AngTwist0Phi}) does not act on the RN metric.
In this way we ensure that the geometry remains intact by the deformation. For details of the construction the reader is referred to references \cite{Ciric:2017rnf},\cite{Ciric:2019uab}.


In index notation, the action  (\ref{NCActionGeometric}) can be written in the form
\begin{eqnarray}
S[\hat{\phi}, \hat{A}] &=& S_\phi + S_A, \nn\\ 
S_\phi &=& \int \d ^4x \, \sqrt{-g}\star\Big( g^{\mu\nu}\star D_{\mu}\hat{\phi}^+ \star
D_{\nu}\hat{\phi} - \mu^2\hat{\phi}^+ \star\hat{\phi}\Big) , \label{SPhi}\\
S_A &=& -\frac{1}{4q^2} \int \d^4 x\,\sqrt{-g}\star g^{\alpha\beta}\star g^{\mu\nu}\star
\hat{F}_{\alpha\mu}\star \hat{F}_{\beta\nu} .\label{SA}
\end{eqnarray}
where the covariant derivative is defined as
\begin{equation}
D_\mu\hat{\phi} = \partial_\mu\hat{\phi} - i \hat{A}_\mu\star \hat{\phi} \label{DPhi}. \nn
\end{equation}

The list of symmetries of the action  (\ref{NCActionGeometric}) (or equivalently of the action (\ref{SPhi}),(\ref{SA}))
can be supplemented by the following infinitesimal $U(1)_\star$
gauge transformations,
\begin{eqnarray}
\delta^\star \hat{\phi} &=& i\hat{\Lambda} \star \hat{\phi}, \nn\\
\delta^\star \hat{A}_\mu &=& \partial_\mu\hat{\Lambda} + i[\hat{\Lambda} \ds \hat{A}_\mu],
\label{NCGaugeTransf}\\
\delta_\star \hat{F}_{\mu\nu} &=& i[\hat{\Lambda} \ds \hat{F}_{\mu\nu}],\nn\\
\delta^\star g_{\mu\nu} &=& 0.\nn
\end{eqnarray}
where $\hat{\Lambda}$ is the NC gauge parameter.
The last transformation in (\ref{NCGaugeTransf}) makes clear that the model (\ref{NCActionGeometric}) studied here is semiclassical. By this we mean that only scalar and gauge field are subject to a NC deformation, while gravitational field remains  unaffected. In this sense, the gravitational field as described in the  model (\ref{NCActionGeometric}) may be deemed classical. The most general situation dealing with noncommutative deformation of gravity becomes  increasingly more involved. For more details see \cite{EPL2017},\cite{PL09},\cite{PLSWGeneral} and references therein.


Our approach is perturbative and we have to expand the action (\ref{SPhi}) up to first order in the deforamtion parameter $a$. To do that we expand the $\star$-products in (\ref{SPhi}) and use the Seiberg-Witten (SW) map. SW map enables to express NC variables as functions of the coresponding commutative variables. In this
way, the problem of charge quantization in $U(1)_\star$ gauge theory does not exist. In the case of
NC Yang-Mills gauge theories, SW map guarantees that the number of degrees of freedom in the NC
theory is the same as in the corresponding commutative theory. That is, no new degrees of freedom
are introduced.

Using the SW-map NC fields can be expressed as function of corresponding commutative fields
and can be expanded in orders of the deformation parameter $a$. Expansions for an arbitrary Abelian
twist deformation are known to all orders \cite{PLSWGeneral}. Applying these results to the twist
(\ref{AngTwist0Phi}), 
expansions of fields up to first
order in the deformation parameter $a$ follow. They are given by:
\begin{eqnarray}
\hat{\phi} &=& \phi -\frac{1}{4}\theta^{\rho\sigma}A_\rho(\partial_\sigma\phi + D_\sigma
\phi), \label{HatPhi}\\
\hat{A}_\mu &=& A_\mu -\frac{1}{2}\theta^{\rho\sigma}A_\rho(\partial_\sigma A_{\mu} +
F_{\sigma\mu}). \label{HatA}  \\
\hat{F}_{\mu\nu} &=& F_{\mu\nu} - \frac{1}{2}\theta^{\rho\sigma}A_{\rho}(\partial_\sigma F_{\mu\nu}
+ D_\sigma F_{\mu\nu})+\theta^{\rho\sigma}F_{\rho\mu}F_{\sigma\nu}. \label{HatFmunu}
\end{eqnarray}
The $U(1)$ covariant derivative of $\phi$ is defined as $D_\mu \phi = (\partial_\mu - i A_\mu)
\phi$. Using the SW-map solutions and expanding the $\star$-products in (\ref{SPhi}) we
find the action up to first order in the deformation parameter $a$. It is given by
\begin{eqnarray}
S_\phi + S_A &=& \int
\d^4x\sqrt{-g}\,
\Big( g^{\mu\nu}D_\mu\phi^+D_\nu\phi -\mu^2\phi^+\phi \label{SExp}\\
&& + \frac{\theta^{\alpha\beta}}{2}g^{\mu\nu}\big( -\frac{1}{2}D_\mu\phi^+F_{\alpha\beta}
D_\nu\phi
+(D_\mu\phi^+)F_{\alpha\nu}D_\beta\phi + (D_\beta\phi^+)F_{\alpha\mu}D_\nu\phi\big) \Big)
.\nn 
\end{eqnarray}
In  (\ref{SExp}) the coupling constant $q$ is absorbed into a definition of $A_\mu$, so that $A_\mu \rightarrow
qA_\mu$.

 Since the RN background also fixes the electromagnetic setting, we may write for the $U(1)$ gauge field $A_\mu$ and the corresponding field strength tensor $F_{\mu\nu},$
\begin{equation}
A_0 = -\frac{qQ}{r}, \qquad  F_{r0} = \frac{qQ}{r^2}.   \label{A0}
\end{equation}
Varying the action (\ref{SExp}) with respect to the field $\phi$ and taking into account (\ref{A0}), one gets
the equation of motion in the form
\begin{eqnarray}
&&
\Big( \frac{1}{f}\partial^2_t -\Delta + (1-f)\partial_r^2 
+\frac{2MG}{r^2}\partial_r + 2iqQ\frac{1}{rf}\partial_t -\frac{q^2Q^2}{r^2f}\Big)\phi
\nonumber\\
&& +\frac{aqQ}{r^3}
\Big( (\frac{MG}{r}-\frac{GQ^2}{r^2})\partial_\varphi
+ rf\partial_r\partial_\varphi \Big) \phi =0, \label{EomPhiExp1}
\end{eqnarray}
where $\Delta$ is the usual Laplace operator. 

In order to solve this equation one assumes \cite{Ciric:2017rnf} the ansatz 
\begin{equation}
\phi_{lm}(t,r,\theta,\varphi) = R_{lm}(r)e^{-i\omega t}Y_l^m(\theta, \varphi) \label{AnsatzPhi} 
\end{equation}
with spherical harmonics $Y_l^m(\theta, \varphi)$. Inserting (\ref{AnsatzPhi}) into
(\ref{EomPhiExp1}) yields an equation for the radial function $R_{lm}(r)$
\begin{eqnarray}
&& f R_{lm}'' + \frac{2}{r}\big( 1-\frac{MG}{r}\big) R_{lm}' - \Big( \frac{l(l+1)}{r^2} 
- \frac{1}{f}(\omega - \frac{qQ}{r})^2   + \mu^2  \Big)R_{lm} \nn\\
&& -ima\frac{qQ}{r^3}\Big( (\frac{MG}{r}-\frac{GQ^2}{r^2})R_{lm} + rf R_{lm}' \Big) =0
.\label{EoMR} 
\end{eqnarray}
While the first line of this equation describes the system without deformation, the second line in  its entirety arises from a quantum  deformation of spacetime. From now on we set $G = 1$.


\section{Results for the QNM spectrum}

In order to obtain a QNM spectrum for the massive scalar perturbations of the  background RN metric (\ref{dsRN}),  under conditions in which the underlying spacetime structure is deformed,
we need to solve the equation (\ref{EoMR}) under specific boundary conditions, those that require purely ingoing waves at the horizon and purely outgoing waves at infinity.
The asymptotic form of the soultions that are fully congruent with QNM boundary conditions may be expressed in terms of the tortoise coordinate $y$ \cite{Ciric:2019uab} in a manner as
\begin{gather}    
R(r) \rightarrow 
\begin{cases}         
Z_{out} e^{i \Omega y} y^{-1-i \frac{\omega qQ - \mu^2  M}{\Omega} - amqQ \Omega}, & \text{for } r \rightarrow \infty,\>\> (y\rightarrow \infty) \\  
  \\
Z_{in}  e^{-i \Big( \omega  - \frac{ qQ}{r_+} \Big)  \Big( 1 + iam  \frac{ qQ}{r_+}  \Big)y },   & \text{for }r \rightarrow  r_+, \>\> (y\rightarrow -\infty)                
\end{cases} . \label{ncboundaryconditions}   
\end{gather}
where $Z_{out}$ and  $Z_{in}$ are the respective amplitudes  which do not depend on $r$ (or $y$). 

We shall apply two approaches, the first one is based on the method of continued fractions and the second approach is  essentially an analytic one. This analytic approach is though applicable to only a near extremal regime of  the system parameters. However, since  the near extremal regime is where a convergence of the continued fraction method  appears to be rather slow, and hence this method  does not seem to be the most adequate  in that case, the latter approach based on analytic method may be considered  as complemental. In this sense both these methods complement each other and  when applied together, they  form a coherent combination, covering results for all possible ranges of the system parameters. The only exception may be the exactly extremal case which might need a specific modification of the continued fraction method \cite{Richartz:2015saa}.

\subsection{Method of continued fractions}

Therefore, in order  to determine
the QNM spectrum of a massive charged scalar field around the RN black hole in the
presence of the noncommutative deformation of space-time, we first implement the continued fraction method \cite{gautschi},\cite{leaver},\cite{nollert}.
The same method was used in \cite{Konoplya:2013rxa},\cite{QNMRNBrazilci},\cite{Chowdhury:2018izv}  to study  the undeformed (commutative) (un)charged scalar and
Dirac QNM spectrum in the RN background.

As the equation (\ref{EoMR}) has an irregular singularity at $r=+\infty$ and three regular singularities at
$r=0$, $r=r_-$ and $r=r_+$, the implementation of Leaver's method    assumes writing  the solution in terms of powers series
around  $r = r_+$. Then the radial part of the scalar field looks as
\begin{equation}  \label{generalpowersolution}
R(r) = e^{i \Omega r}  {(r-r_-)}^{\epsilon} \sum_{n=0}^{\infty} a_n {\Big( \frac{r-r_+}{r-r_-} \Big)}^{n + \delta},
\end{equation}
 where, in accordance with (\ref{ncboundaryconditions}), the parameters $\Omega,$ $\delta$  and  $\epsilon$ are fixed to be
\begin{equation}  \label{epsilondelta}
\Omega^2 = \omega^2 - \mu^2, \qquad
\delta = -i \frac{r_+^2}{r_+ - r_-} \Big( \omega - \frac{qQ}{r_+} \Big), \qquad
\epsilon = -1 - i qQ \frac{\omega}{\Omega} + i \frac{r_+ + r_-}{2\Omega} \Big(  \Omega^2 + \omega^2  \Big).
\end{equation}

 Inserting the general expansion
(\ref{generalpowersolution}) into the equation (\ref{EoMR}) gives rise to  the following $6$-term recurrence relations
\begin{eqnarray}  \label{contfr}
  A_n a_{n+1} + B_n a_n +C_n a_{n-1} + D_n a_{n-2} + E_n a_{n-3} + F_n a_{n-4 }  = 0, \nonumber \\
   A_3 a_{4} + B_3 a_3 +C_3 a_{2} + D_3 a_{1} + E_3 a_{0}   = 0, \nonumber \\
  A_2 a_{3} + B_2 a_2 +C_2 a_{1} + D_2 a_{0}   = 0, \nonumber \\
  A_1 a_{2} + B_1 a_1 +C_1 a_{0}    = 0, \nonumber \\
   A_0 a_{1} + B_0 a_0   = 0, 
\end{eqnarray}
where the coefficients $A_n, B_n, C_n, D_n, E_n $ and $ F_n$ are given as
\begin{equation}   \label{contfr1}
\begin{split}
 & A_n   =   r_+^3 \alpha_{n}, \nonumber \\
 &  B_n  =  r_+^3 \beta_n -iamqQ (r_+ - r_-) r_+ (n + \delta) - \frac{1}{2} iamqQ (r_+ + r_-) r_+   \nonumber \\
  &  \qquad +  iam qQ r_+ r_- - 3 r_+^2 r_- \alpha_{n-1}, \nonumber \\
 & C_n  =  r_+^3 \gamma_n  + 3r_+ r_-^2 \alpha_{n-2}    +iamqQ (r_+ - r_-)(2 r_+ +r_-) (n + \delta -1) -
    iamqQ (r_+ - r_-)r_+ \epsilon   \nonumber \\ 
 & \qquad +  \frac{1}{2}  iamqQ (r_+ + r_-)(2r_+ + r_-) - 3iamqQ r_+ r_- + amqQ \Omega {(r_+ - r_-)}^2  r_+  -3r_+^2 r_- \beta_{n-1} + ,  \nonumber \\
& D_n  = - r_-^3 \alpha_{n-3}  + 3r_+ r_-^2 \beta_{n-2} -3 r_+^2 r_- \gamma_{n-1} +iamqQ (r_+^2 - r_-^2)\epsilon
  + 3iam qQ r_+ r_- 
  -amqQ \Omega {(r_+ - r_-)}^2 r_-   \nonumber \\
&  \qquad   -iamqQ (r_+ - r_-) ( r_+ +2r_-) (n + \delta -2) -  \frac{1}{2}  iamqQ (r_+ + r_-)(r_+ +2 r_-) ,  \nonumber \\
 &  E_n  =   3r_+ r_-^2 \gamma_{n-2} - r_-^3 \beta_{n-3} + iamqQ (r_+ - r_-) r_-  (n + \delta -3) \nonumber \\
 &  \qquad     -iamqQ (r_+ - r_-)r_- \epsilon
     +  \frac{1}{2}  iamqQ (r_+ + r_-)r_- iam qQ r_+ r_- ,  \nonumber \\
 &  F_n = -r_-^3 \gamma_{n-3},     
\end{split}
\end{equation}
and the coefficients $\alpha_n, \beta_n, \gamma_n$  are  the analogous coefficients appearing in the $3$-term recurrence relations which result from (\ref{EoMR}) when the deformation $a$ is switched off (see the references \cite{Ciric:2019uab},\cite{QNMRNBrazilci},\cite{Chowdhury:2018izv}).
The first relation in  (\ref{contfr})  is a general $6$-term recurrence relation, while the remaining four relations are the indicial equations relating the lowest order coefficients $a_n$ in the general expansion (\ref{generalpowersolution}).
They serve as boundary conditions for the  first relation in  (\ref{contfr}).

The $6$-term recurrence relations need to be reduced to the  $3$-term recurrence relations
\begin{eqnarray}    \label{contfrsimplegauss}
  A_n^{(3)} a_{n+1} + B_n^{(3)} a_n +  C_n^{(3)} a_{n-1} = 0, \nonumber \\
   A_0^{(3)} a_{1} + B_0^{(3)} a_0   = 0.  
\end{eqnarray}
This is achieved in a gradual process which consists of applying the Gauss elimination successively three times in a row (for more details see the reference \cite{Ciric:2019uab}).
As a result of this process, one gets  the coefficients of the third level, $A_n^{(3)}, B_n^{(3)}, C_n^{(3)} $ expressed  in terms of both  the coefficients of  the zeroth level $A_n, B_n, C_n, D_n, E_n, F_n$ and the coefficients $\alpha_n, \beta_n, \gamma_n$.
Having that, the QNM frequencies, and in particular  the fundamental QNM frequency will be obtained by solving the equation
\begin{equation}   \label{infcontfracgauss}
0= B_0^{(3)} - \cfrac{A_0^{(3)} C_1^{(3)}}{B_1^{(3)} - \cfrac{A_1^{(3)} C_2^{(3)}}{B_2^{(3)} -\cfrac{A_2^{(3)} C_3^{(3)}}{ B_3^{(3)} - \cdot \cdot \cdot  \cfrac{A_n^{(3)} C_{n+1}^{(3)}}{B_{n+1}^{(3)} - \cdot \cdot \cdot}  }}}.
\end{equation}

Using a specially devised  root finding algorithm \cite{Ciric:2019uab} as applied to (\ref{infcontfracgauss}) to determine the QNM spectrum, it is possible to portray graphically a behaviour of the fundamental mode as a function of the electromagnetic coupling $qQ$ (Figures 1 and 2), as well as to portray its behaviour as a function of the scalar field mass $\mu$ (Figures 3 and 4).
 In both cases the NC parameter $a$ is set to $a=0.01$. 

In effect Figures 1 and 2 show respectively
a dependence of the real and imaginary part of the fundamental QNM frequency $\omega = {\mbox Re}\, \omega +i {\mbox Im}\,
\omega$ on the charge of the scalar field
$qQ$, whereat the charge of the RN black hole $Q$ is kept fixed. Other  parameters in the model are given as: $\mu =0.05$, $l=1$ and  $M=1$. The ratio $\frac{Q}{M}$ between the charge and the mass of the RN black hole measures how much this black hole is close to the extremal condition. Figures 1 and 2 present the results for
ten different values of  the ratio $\frac{Q}{M}$, which runs in between $0.01$ and $0.999999$.

Likewise,  Figures 3 and  4  show a dependence of the fundamental frequency Re\,$\omega$  and Im\,$\omega$ on
the mass $\mu$ of the scalar field $\phi$. Other system parameters are given as:
$qQ=1$, $l=1$ and $M=1$. The results for different ratios  $\frac{Q}{M}$  are shown in different colors.
The values of $\frac{Q}{M}$ run in between $0.01$ and $0.9999$.

The characteristic features of the fundamental mode that can be drawn from Figures 1 and 2 include a mostly linear behaviour of the real part of the fundamental frequency with respect to the charge $q$ (at least for the larger values of $qQ$) and the existence of a specific constant value at which the imaginary part of the fundamental frequency saturates. Moreover, it is clearly visible that there exists a critical value of the electromagnetic
coupling at which the real part of the fundamental frequency Re\,$\omega$ approaches zero as $qQ$  decreases.
 It is also interesting to observe a behaviour of the fundamental mode as the extremal limit is approached, $\frac{Q}{M} \rightarrow 1.$ We can see from Figures 1 and 2 that in this limit the real part of the fundamental frequency behaves as  $\frac{qQ}{r_+} \approx q$, while the imaginary part acquires  smaller and smaller values.

At the same time, the characteristic feature of the fundamental mode that is clearly visible from Figures 3 and 4
includes  the appearance of quasi-resonances, special values of the mass when ${\mbox Im}\,\omega =0$, implying the existence of non-decaying modes.
They are present  for all values of the ratios  $\frac{Q}{M}$.


\begin{figure}[h]
\begin{minipage}{20pc}
\includegraphics[width=20pc]{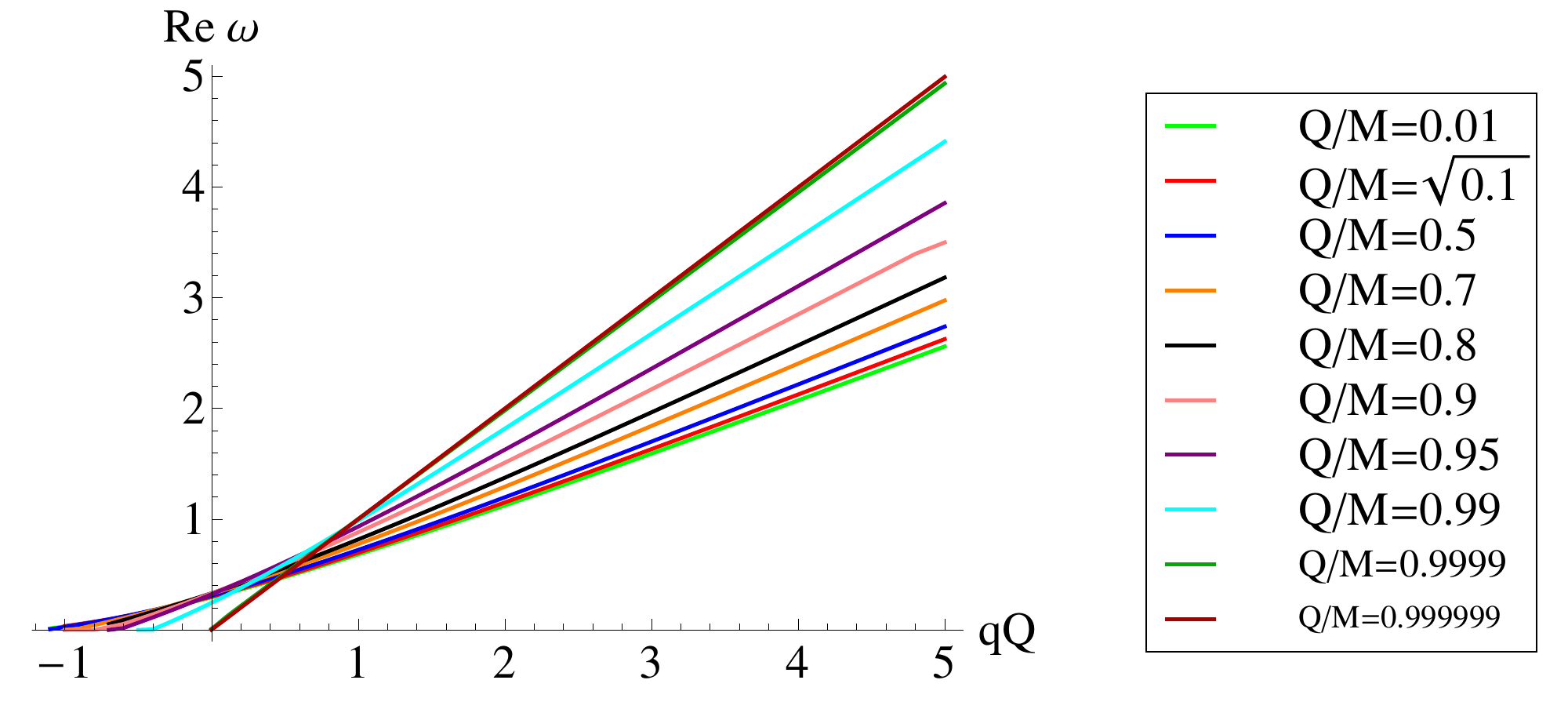}
\caption{\label{label}    Dependence of Re\,$\omega$  on the charge $qQ$ of the scalar field with the mass  $\mu=0.05$ and the orbital momentum $l=1$. Different extremalities are shown in different colors.}
\end{minipage}\hspace{2pc}%
\begin{minipage}{20pc}
\includegraphics[width=20pc]{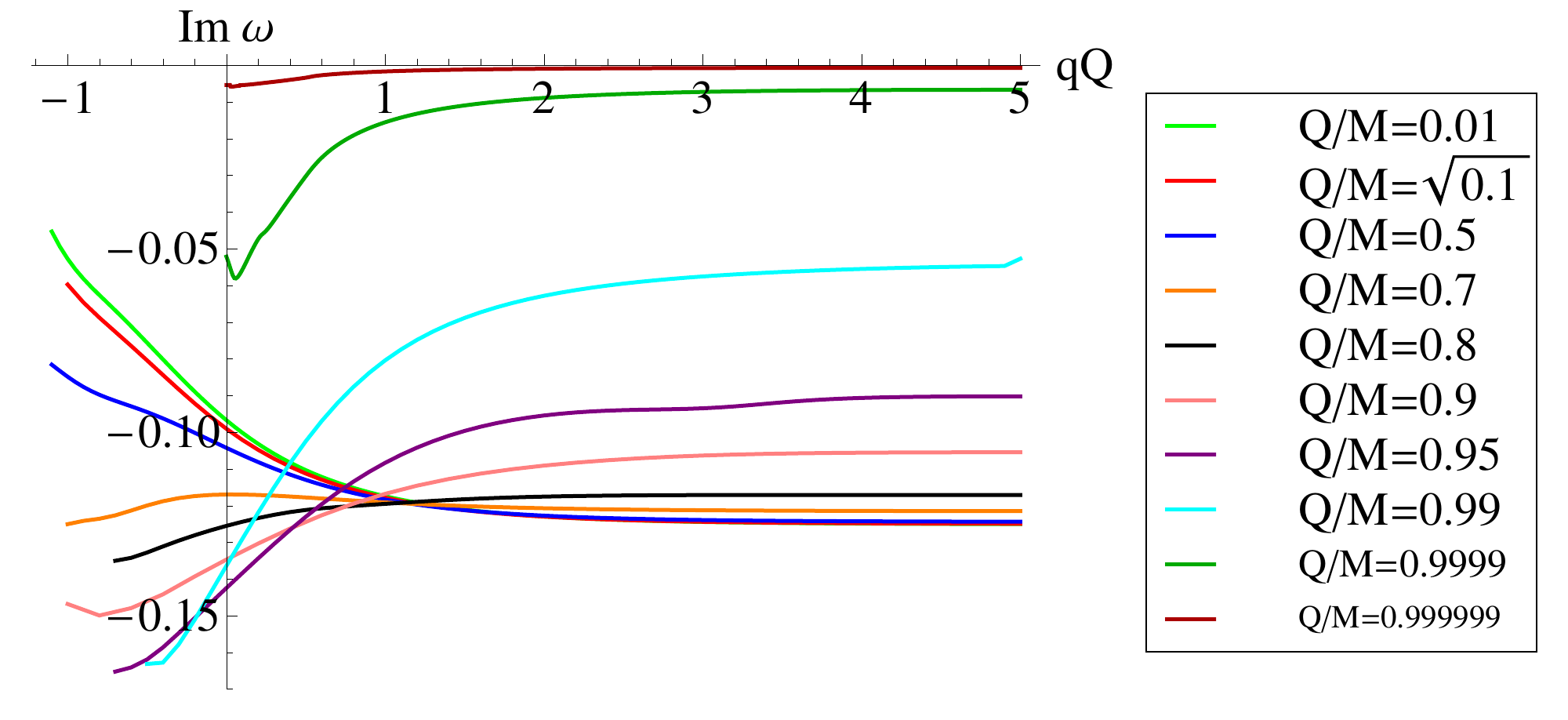}
\caption{\label{label}  Dependence of  Im\,$\omega$  on the charge $qQ$ of the scalar field with the mass  $\mu=0.05$ and the orbital momentum $l=1$. Different extremalities are shown in different colors.}
\end{minipage} 
\end{figure}

\begin{figure}[h]
\begin{minipage}{20pc}
\includegraphics[width=20pc]{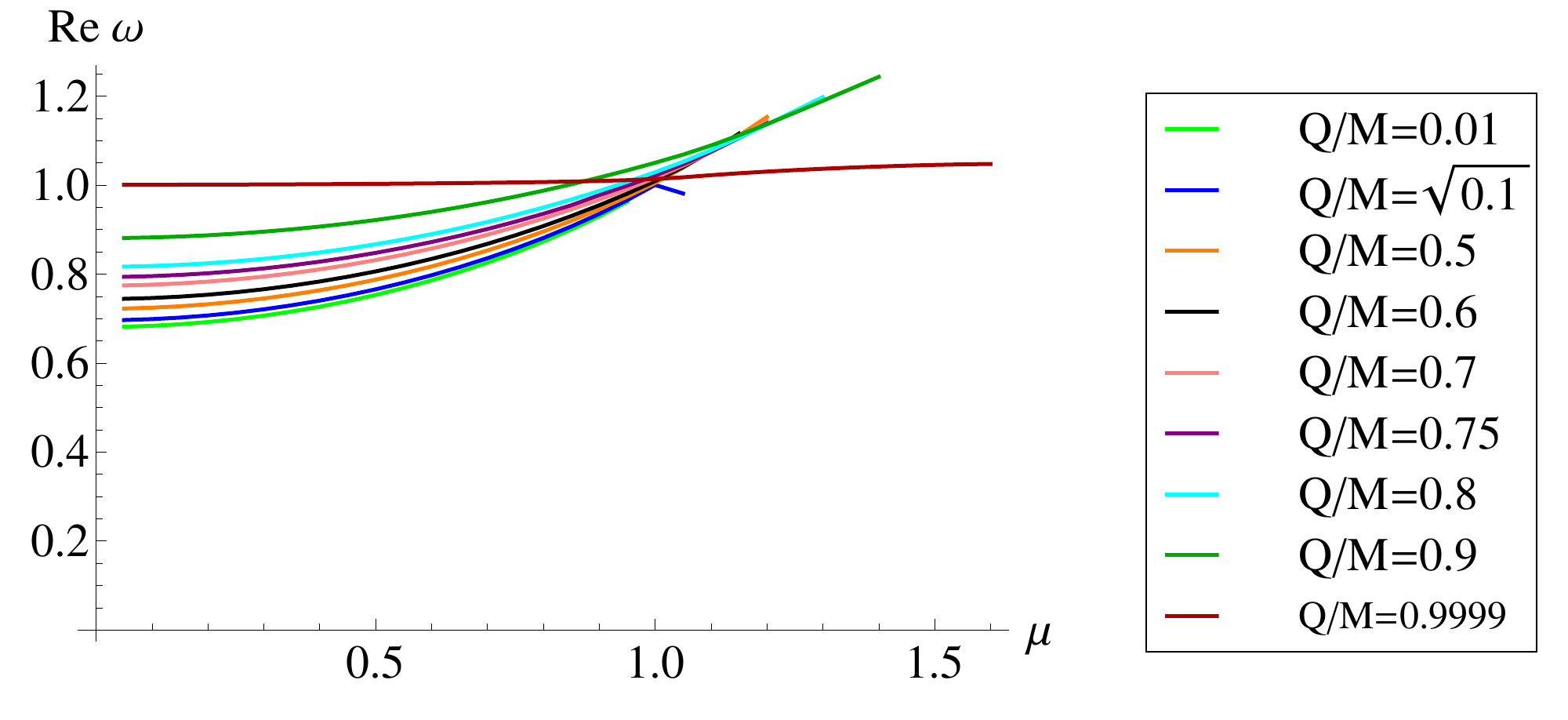}
\caption{\label{label}     Dependence of Re\,$\omega$  on the mass $\mu$ of the scalar field with the charge $qQ=1$ and the orbital momentum $l=1$.}
\end{minipage}\hspace{2pc}%
\begin{minipage}{20pc}
\includegraphics[width=20pc]{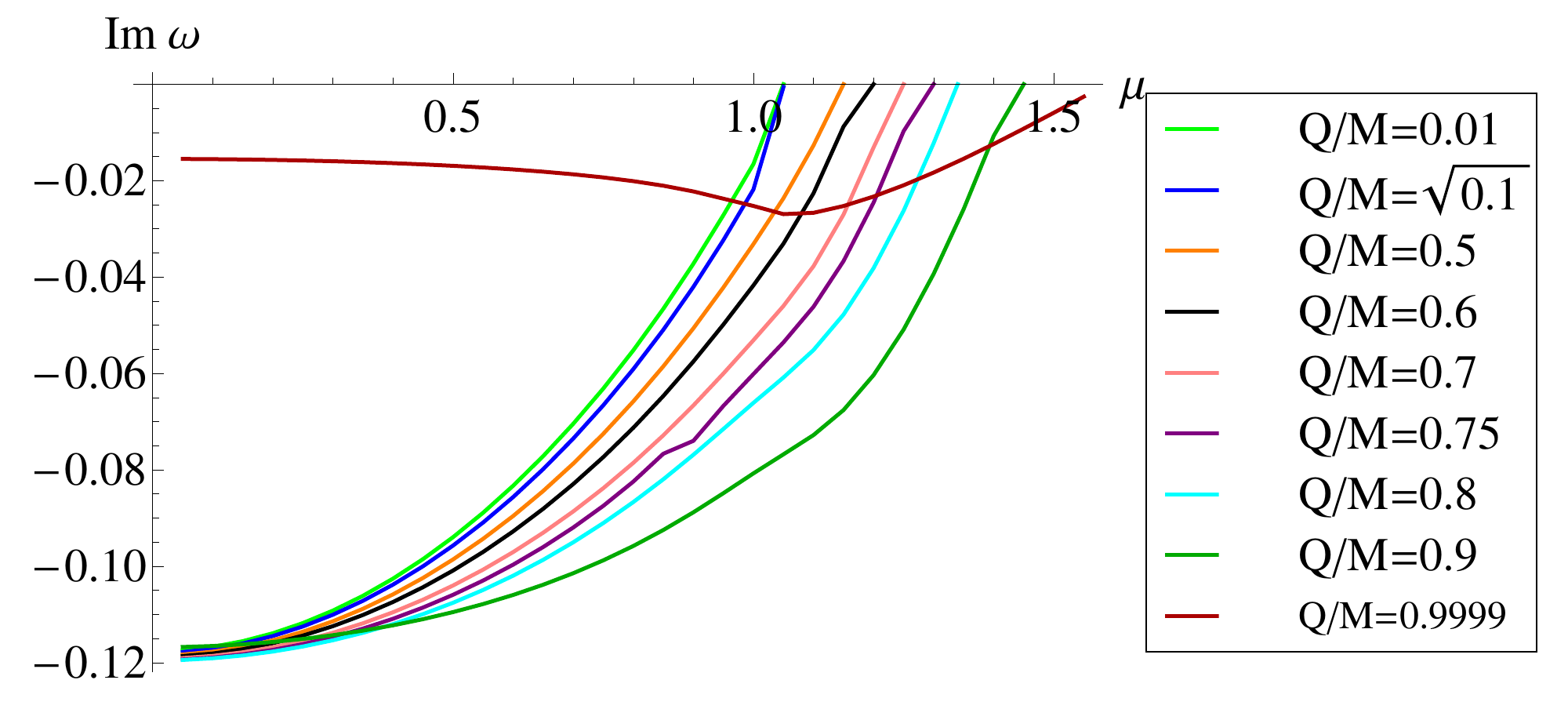}
\caption{\label{label}   Dependence of  Im\,$\omega$  on the mass $\mu$ of the scalar field with the charge $qQ=1$ and the orbital momentum $l=1$.}
\end{minipage} 
\end{figure}

As the effects of noncommutative deformation in our model are not clearly visible from Figures 1-4, we have introduced
a quantity called {\it{frequency splitting}}, which is defined as $\omega^\pm = \omega_{m=\pm 1} -\omega_{m=0},$
where $m = -l,\ -l +1,..., l$ is the projection of the orbital angular momentum $l$.
In case that the orbital angular momentum $l$ is greater than $1,$ the frequency splitting receives a generalization in the form
$\omega^{\pm \pm \cdot \cdot \cdot  \pm} = \omega_m -\omega_{m=0}$. Specifically,  $\omega^{\pm\pm}=\omega_{m=\pm 2}-\omega_{m=0}$. Here we shall consider only quantities 
$\omega^\pm $ and $\omega^{\pm \pm}$. They are demonstrated at Figures 5 and 6 for the  angular momentum channel $l=2$
and for the value of the ratio  $\frac{Q}{M}=0.5$. On these figures the real and the  imaginary part of the frequency splittings
$\omega^\pm $ and $\omega^{\pm \pm}$ is shown as a function of the scalar field charge $q$. 
The frequency splitting can be graphically presented for any value of  the ratio  $\frac{Q}{M}$.

\begin{figure}[h]
\begin{minipage}{20pc}
\includegraphics[width=20pc]{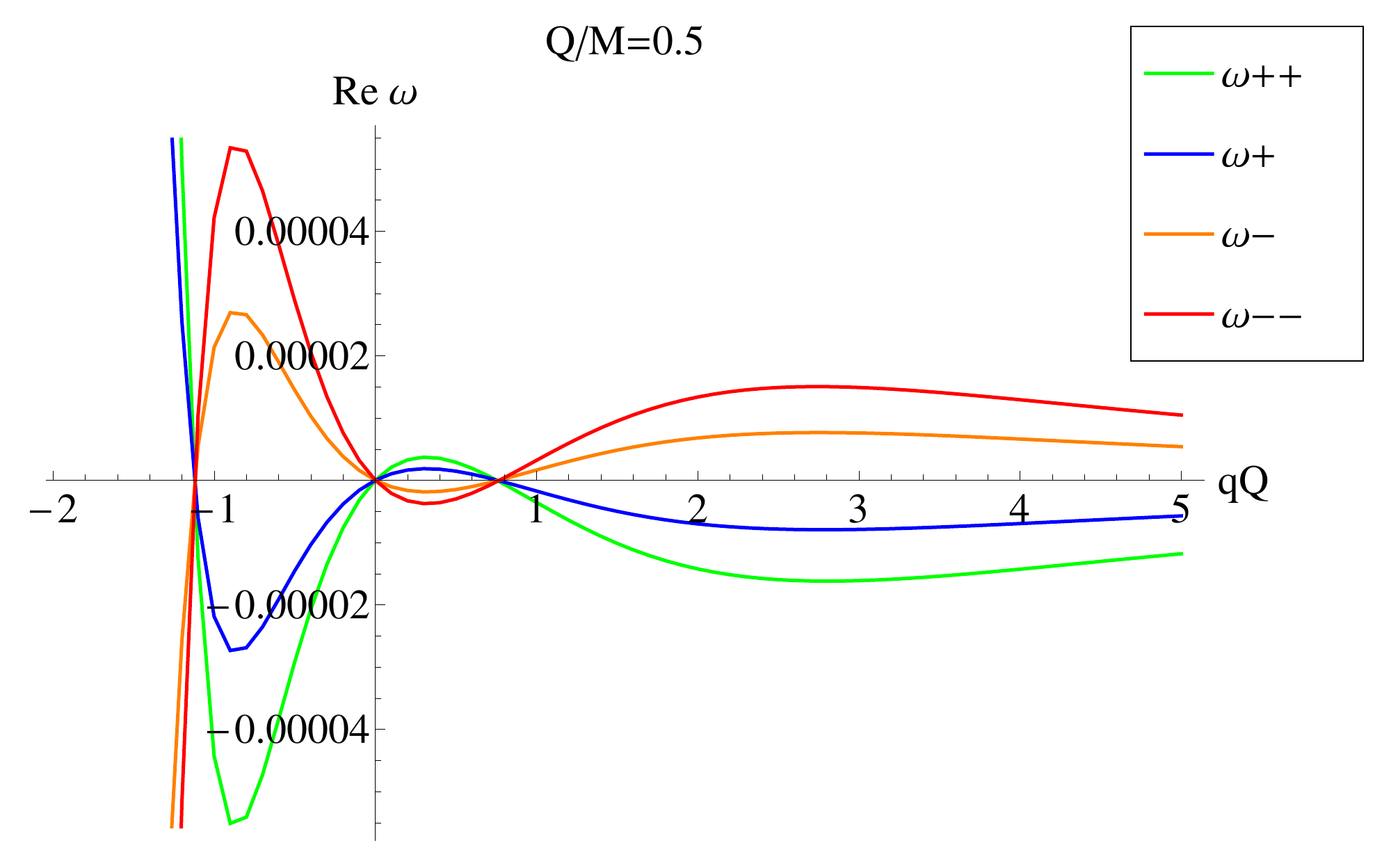}
\caption{\label{label}     Dependence of Re\,$\omega^\pm $, Re\,$\omega^{\pm
\pm} $  on the charge $qQ$ of the scalar field with the mass $\mu=0.05$, orbital momentum $l=2$ and
extremality $\frac{Q}{M}=0.5$.}
\end{minipage}\hspace{2pc}%
\begin{minipage}{20pc}
\includegraphics[width=20pc]{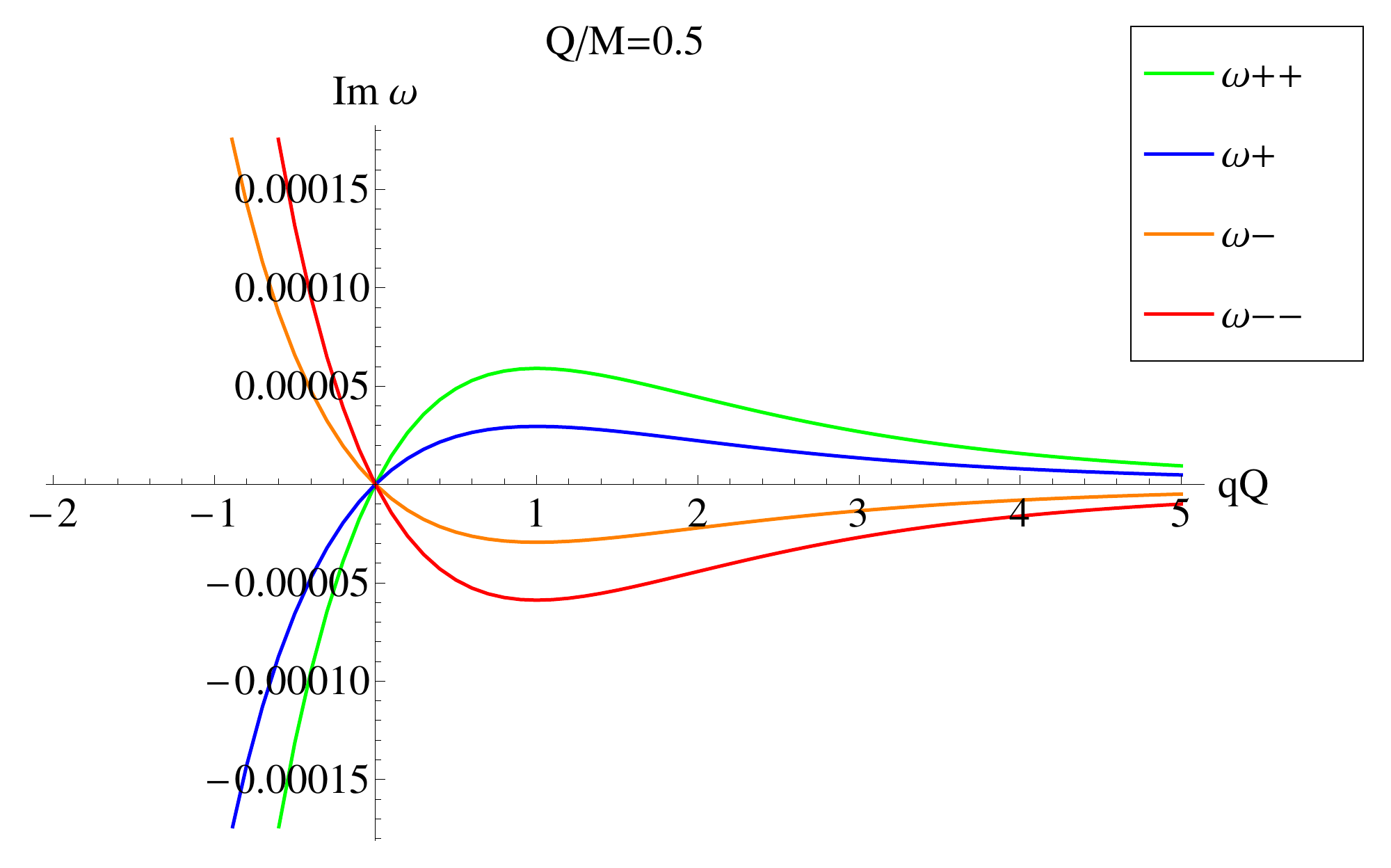}
\caption{\label{label}    Dependence of  Im\,$\omega^\pm$, Im\,$\omega^{\pm \pm}$  on the charge $qQ$ of the scalar field with the mass $\mu=0.05$, orbital momentum $l=2$ and
extremality $\frac{Q}{M}=0.5$.}
\end{minipage} 
\end{figure}


\subsection{Analytic method in the near extremal approximation}

The second approach that we adopt  here is based on an analytic method \cite{Hod:2010hw},\cite{Hod:2017gvn}
 which is however only accurate  in a highly restrictive regime of system parameters 
that is related to conditions of near extremality. The main idea behind this approach is to solve separately the equation  (\ref{EoMR}) in two distinct regions, first in the region relatively close to the horizon and then in the region relatively far from the horizon and afterwards to extrapolate both solutions into a region of common overlap. As a result of  matching of the two extrapolated solutions,
 the  quantization condition emerges
that determines a QNM spectrum for the scalar perturbations of the RN black hole. This quantization condition  is given by
\begin{equation}
\label{QNMcondition}
\begin{split}
&  \frac{\Gamma(1-2i\sigma) \Gamma(-2i\sigma - 2 \tilde{\rho}) }{\Gamma(\frac{1}{2} -i\sigma -ik -
\tilde{\rho} ) \Gamma(\frac{1}{2} -i\sigma +ik - i \Omega -\tilde{\rho} ) \Gamma(\frac{1}{2}
-i\sigma -i \kappa)}   \\
& = - \frac{\Gamma(1+2i\sigma) \Gamma(2i\sigma + 2 \tilde{\rho}) \tau^{-2
\tilde{\rho}}}{\Gamma(\frac{1}{2} +i\sigma -ik + \tilde{\rho} ) \Gamma(\frac{1}{2} +i\sigma +ik - i
\Omega +\tilde{\rho} ) \Gamma(\frac{1}{2} +i\sigma -i \kappa)} \\
& \times {\Big( -2 i \sqrt{\omega^2 -
\mu^2}
~r_+ \tau \Big) }^{ -2i\sigma},
\end{split}
\end{equation}
where the quantities $\tilde{\rho}, k, \sigma, , \kappa, \Omega$ are all functions of the system parameters (the mass and the charge of the RN black hole, as well as the mass and the charge of the scalar field), the frequency $\omega $  and the deformation parameter $a$. In addition, $\tau$ is the extremality parameter, $\tau = \frac{r_+ - r_-}{r_+}.$ For details see the reference \cite{Ciric:2017rnf}.
In general, this condition cannot be solved analytically. In the following, we present
some numerical results for the QNM frequencies, obtained by Wolfram Mathematica which result from
the analytic condition  (\ref{QNMcondition}). In particular, only the fundamental quasinormal mode will be analysed.

In this way Figures 7 and 8 show  a dependence
of the real and imaginary part of the fundamental QNM frequency $\omega$  on the charge  $q$ of the scalar field.
 In order to make comparison for different (albeit still close to $1$) values of the ratio $\frac{Q}{M}$, we plot the results for two
choices of "extremality": $\frac{Q}{M}=0.9999$ and $\frac{Q}{M}=0.999999$. The former one corresponds  to the "extremality" parameter $\tau
=0.0278891617$, while the latter corresponds to $\tau=0.0028244321$. It is important to note that a derivation of  the condition (\ref{QNMcondition}) was made within the approximation that the NC deformation parameter $a$ is of the same order as the extremality $\tau$. Therefore, in the case $\frac{Q}{M}=0.9999$ we set $a=0.1,$ while in the case 
$\frac{Q}{M}=0.999999$ we set $a=0.01$.
The $q$ dependance of $ \mbox{Re}\,\omega$ and
$\mbox{Im}\,\omega$ is presented in Figures 7  and 8, where it was assumed that the mass of the scalar
field is fixed at $\mu=0.05$.

\begin{figure}[h]
\begin{minipage}{20pc}
\includegraphics[width=20pc]{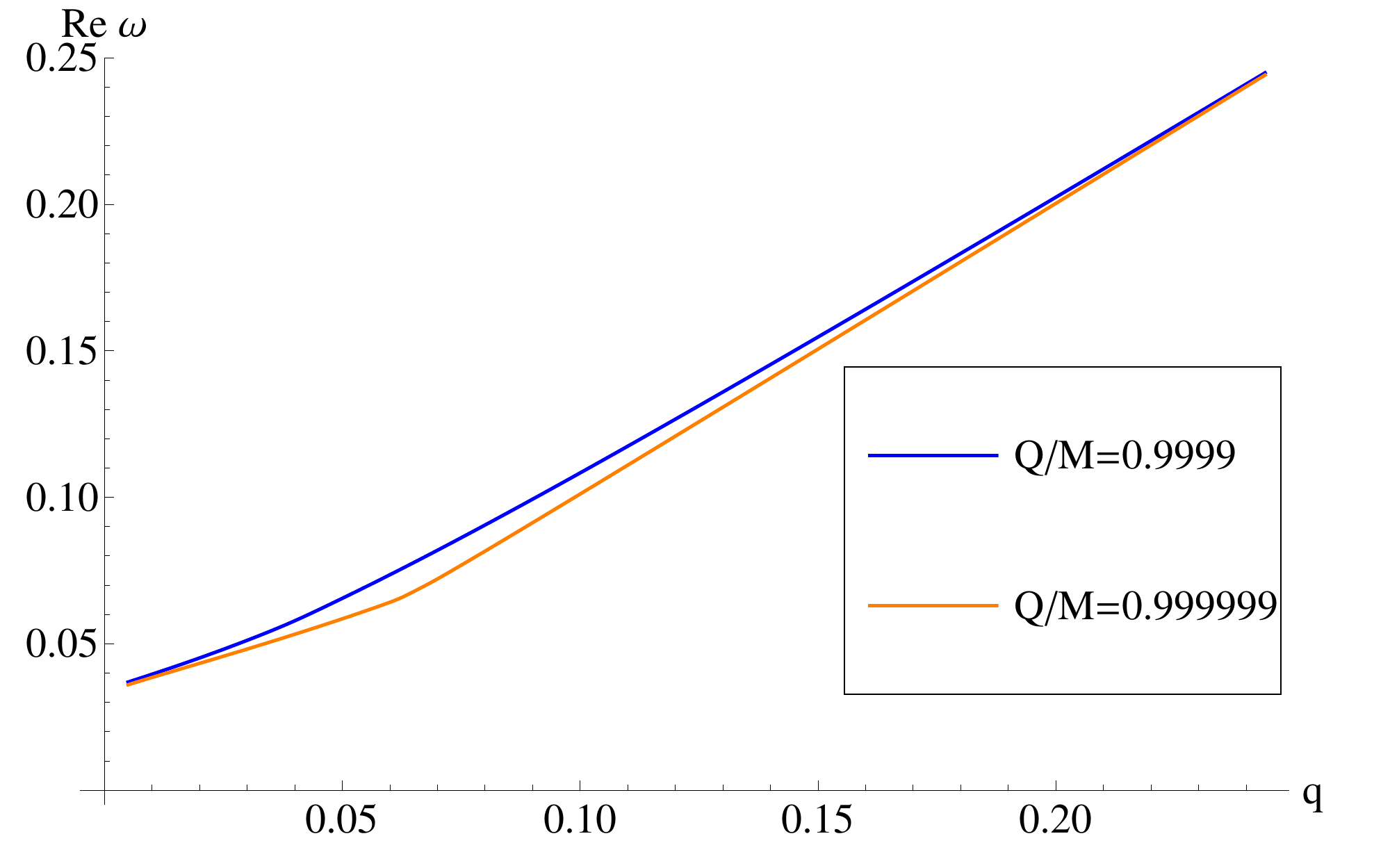}
\caption{\label{label}      Dependance of $ \mbox{Re}\,\omega$ on the charge $q$  of the scalar field with the mass  $\mu=0.05$, $l=1$.} 
\end{minipage}\hspace{2pc}%
\begin{minipage}{20pc}
\includegraphics[width=20pc]{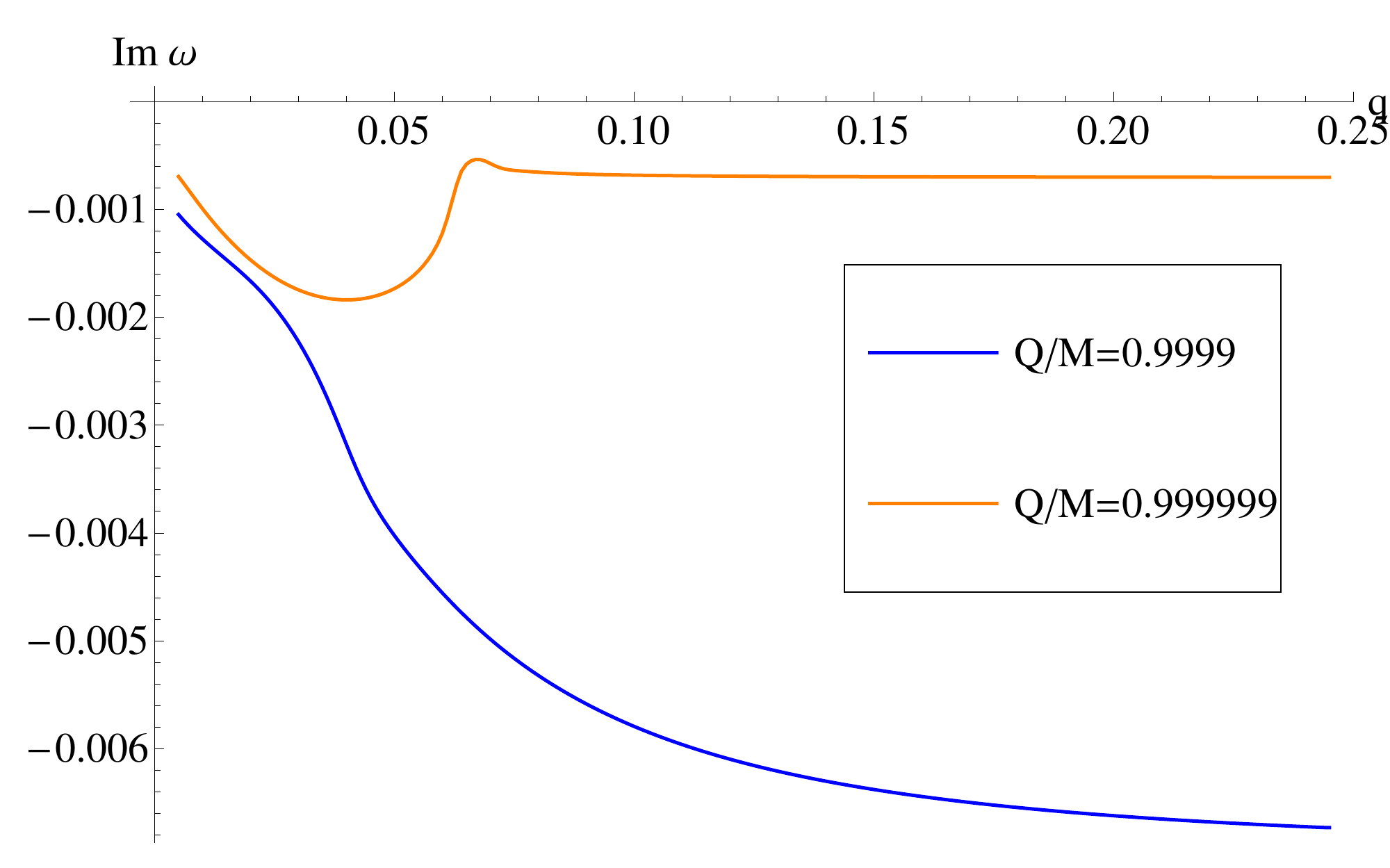}
\caption{\label{label}   Dependance of $ \mbox{Im}\,\omega$ on the charge $q$  of the scalar field with the mass  $\mu=0.05$, $l=1$. }
\end{minipage} 
\end{figure}

Likewise, the dependance of the fundamental QNM frequency $\omega$ on the mass  $\mu$ of the scalar
field, for the  fixed charge  $q=0.075$,  is shown in Figures 9 and 10.

\begin{figure}[h]
\begin{minipage}{20pc}
\includegraphics[width=20pc]{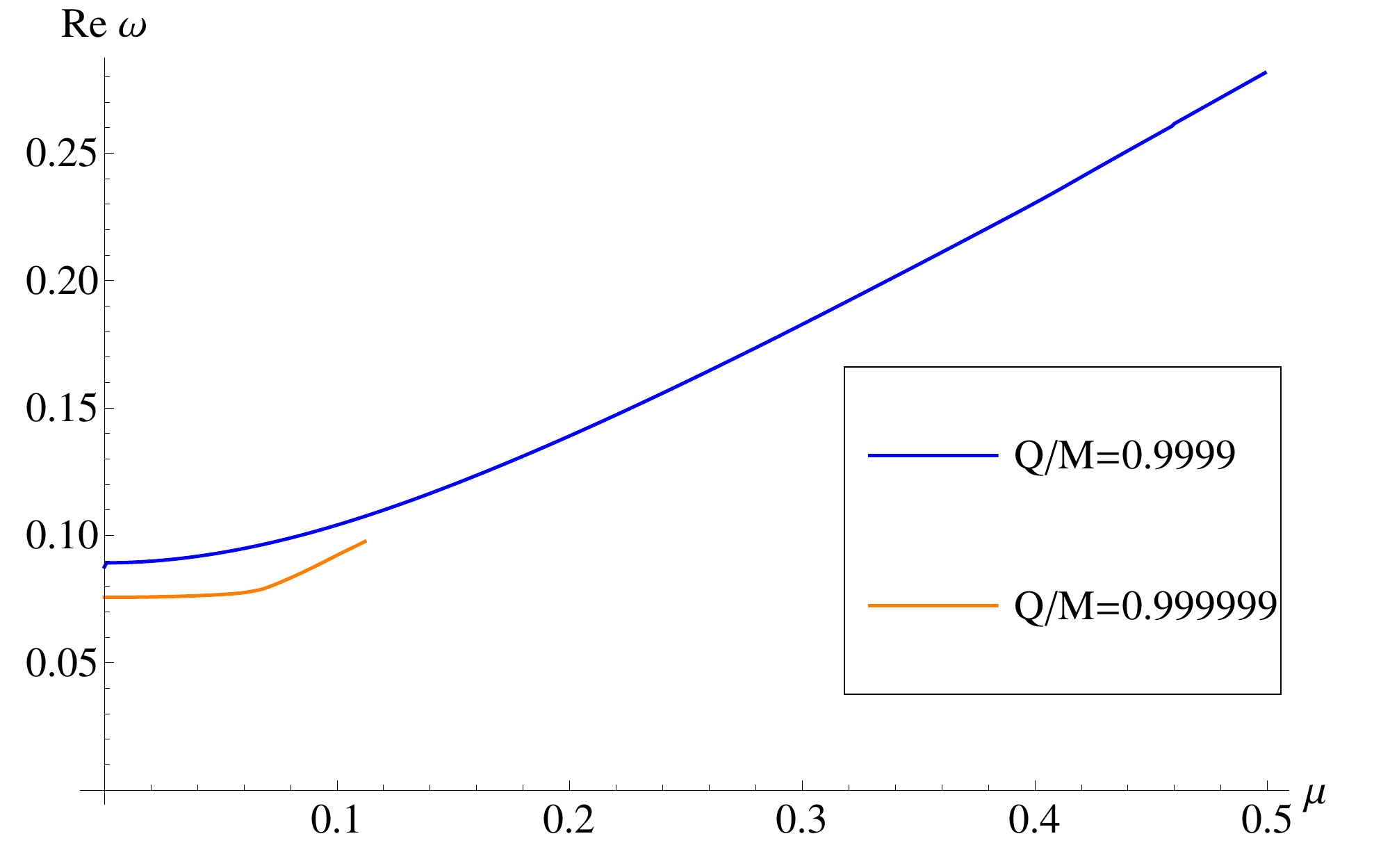}
\caption{\label{label}       Dependance of $ \mbox{Re}\,\omega$ on the mass $\mu$ of the scalar field with the charge $q=0.075$, $l=1$.} 
\end{minipage}\hspace{2pc}%
\begin{minipage}{20pc}
\includegraphics[width=20pc]{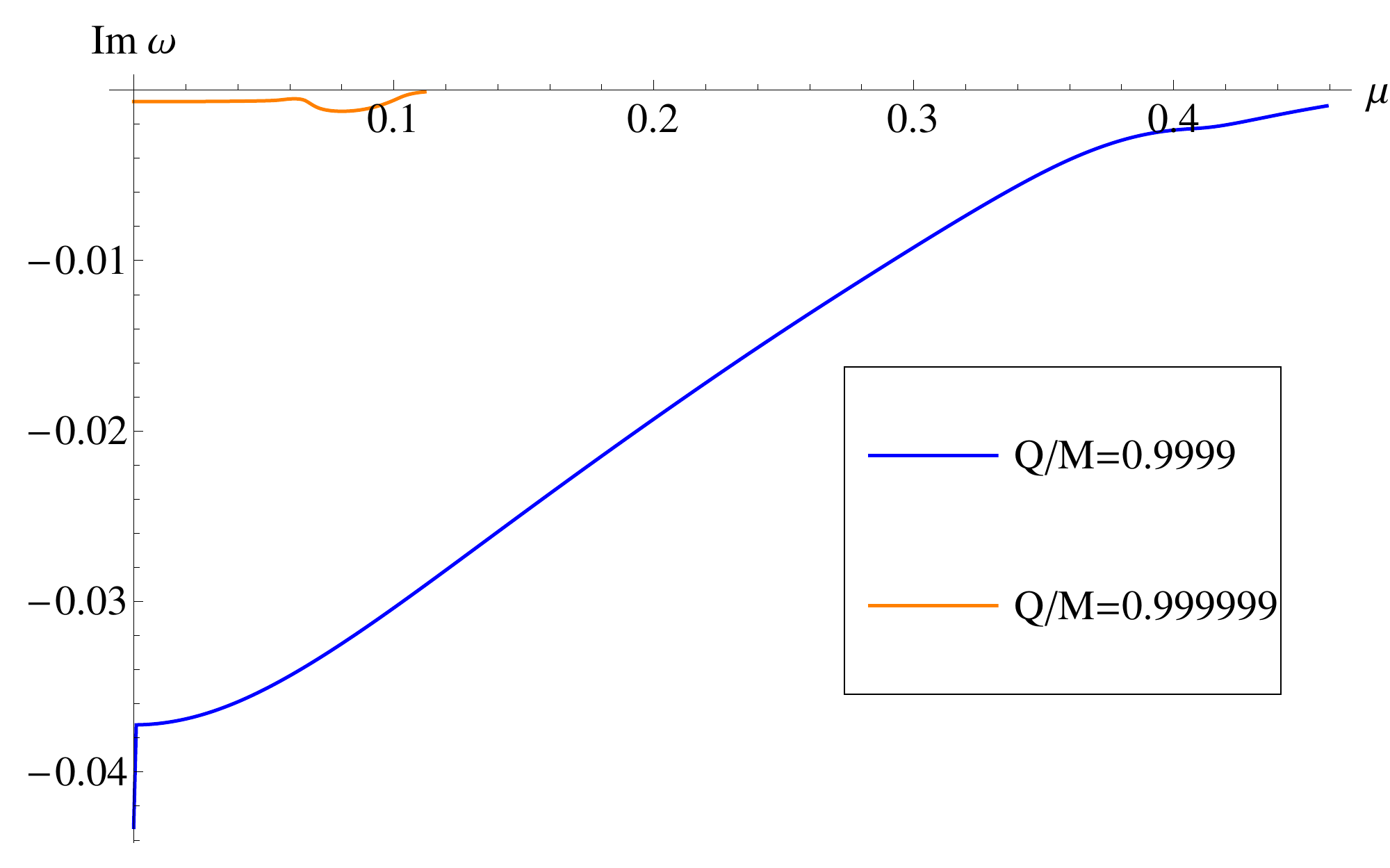}
\caption{\label{label}   Dependance of $ \mbox{Im}\,\omega$ on the mass $\mu$ of the scalar field with the charge $q=0.075$, $l=1$. }
\end{minipage} 
\end{figure}

 As in a previous method, a non-zero NC effect on the QNM spectrum may be inferred by studying the quantities called {\it{frequency splitting}}. They were already defined before and denoted as  $\omega^\pm$, $\omega^{\pm\pm}$.  Figures 11 and 12 show a dependance of  the real  and  imaginary part    of the frequency  splittings $\omega^\pm$, $\omega^{\pm\pm}$  on the scalar field charge $q$. The graphs are made for the following set of parameters: $Q/M=0.999999$, $\mu=0.05$ and $l=2$.

\begin{figure}[h]
\begin{minipage}{20pc}
\includegraphics[width=20pc]{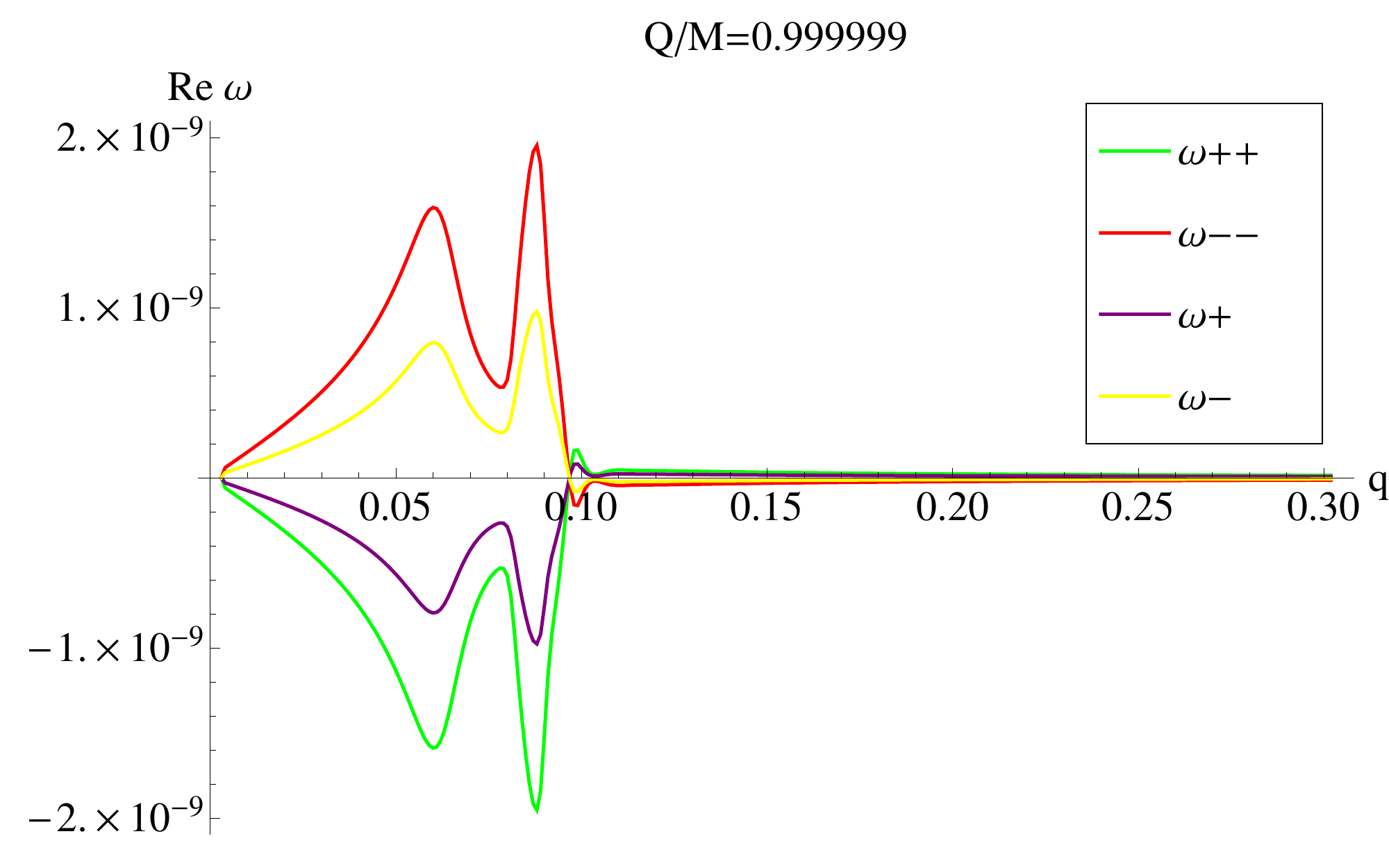}
\caption{\label{label}        Dependance of frequency splitting of  $\mbox{Re}\,\omega$ on the charge $q$ of the scalar field with the mass $\mu=0.05$, $l=2$. } 
\end{minipage}\hspace{2pc}%
\begin{minipage}{20pc}
\includegraphics[width=20pc]{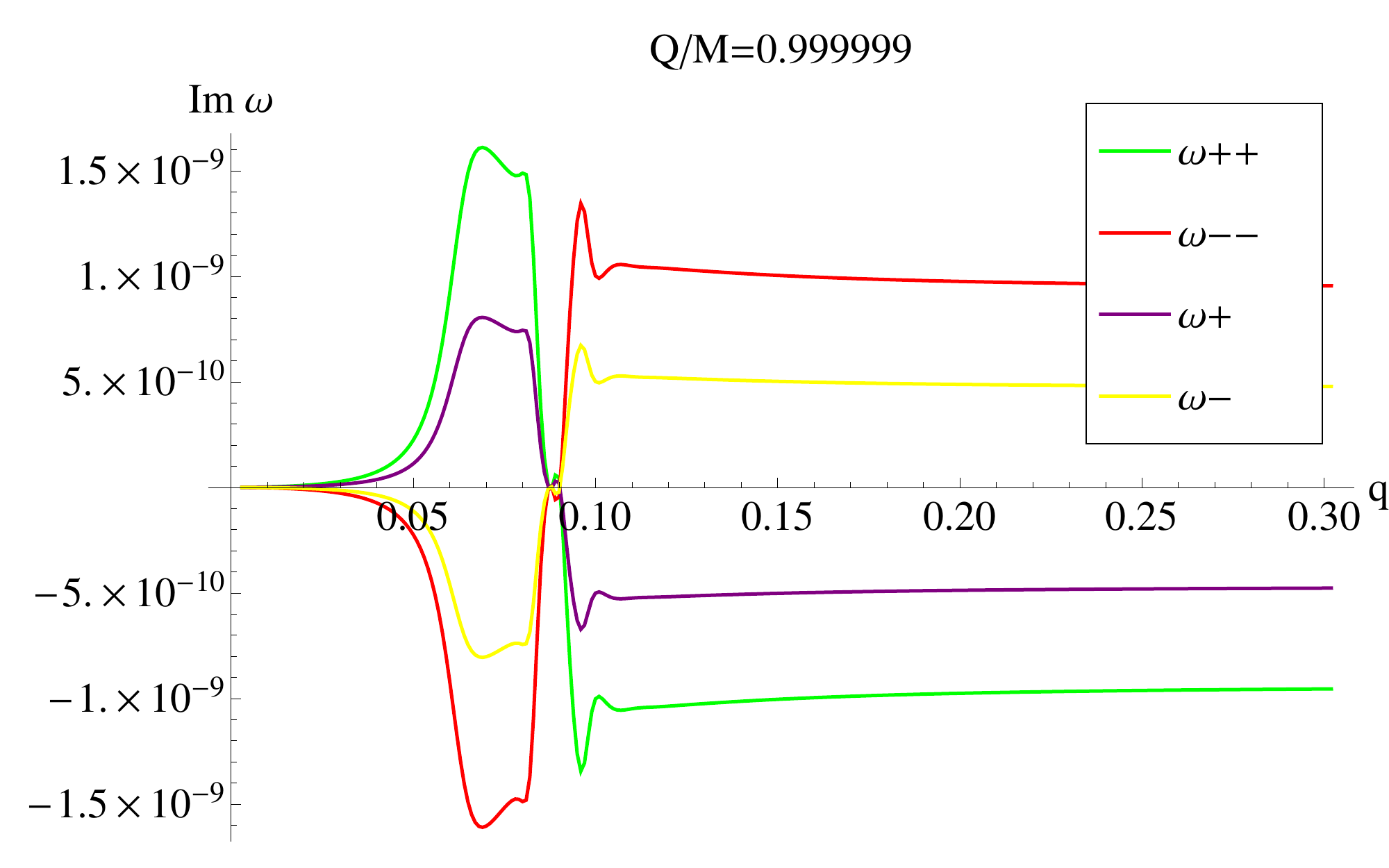}
\caption{\label{label}    Dependance of frequency splitting  of $ \mbox{Im}\,\omega$ on the charge $q$ of the scalar field with the mass $\mu=0.05$, $l=2$. }
\end{minipage} 
\end{figure}

Two latter figures  clearly demonstrate a phenomenon of  frequency splitting, indicating lines that correspond to $m=\pm 2$ and $m=\pm 1$. In this way they show the same patterns already outlined at Figures 5 and 6, thus supporting the findings obtained by the continued fraction method. 
This effect, though being very tiny,  is qualitatively very
important, as it  indicates a presence of  NC driven features in the QNM spectrum, thus signalling an essentially quantum nature of  spacetime.
Since the frequency splitting
has its origin in a coupling between the deformation parameter $a$ and the azimuthal  (magnetic) quantum number $m$ (see equation (\ref{EoMR})),
this  effect  is reminiscent of  a  Zeeman-like splitting 
in the spectrum of the hydrogen-like atoms, with deformation $a$ taking  the role of a magnetic field.

Interestingly, all figures related to frequency splitting (Figures 5,6,11,12) show the same property, namely $~{\omega^{+}} = - {\omega^{-}} ~$ and  $~{\omega^{++}} = - {\omega^{--}}.~$  This feature  was expected due  to a parity symmetry of the model considered.
Other features where two methods agree qualitatively in their findings include: the appearance of
 quasi-resonances in the spectrum,  the existence of constant values at which the imaginary part of the QNM frequency saturates and the appearance of the  linearly dependent pattern for larger values of $qQ$, representing  the real part of the QNM frequency as a function of the  charge $q$ . 

Besides QNM spectra of black holes, there are many other important concepts in physics which may require an appropriate reassessment in terms of  quantum nature of spacetime, such as the geodesic and quantum completeness \cite{Juric:2018qzl}, the existence of naked singularities and related issue of validness of the cosmic censorship hypothesis \cite{Gupta:2019cmo}, the renormalization of black hole entanglement entropy \cite{Juric:2016zey} and a description of physics  in terms of fuzzy (Anti)-de Sitter space \cite{buric},\cite{buric1},\cite{Jurman:2013ota},\cite{Jurman:2017kkp} to name a few. We think that these issues should be given a considerable attention in future studies.

\vskip1cm \noindent 
{\bf Acknowledgement}
The work of M.D.C. and N.K. is   supported by project
ON171031 of the Serbian Ministry of Education and Science. The work of A.S. is  partially supported  by the project  RBI-TWINN-SIN.    This work is partially
supported by ICTP-SEENET-MTP Project NT-03 "Cosmology-Classical and Quantum Challenges" in frame of the Southeastern European Network in  Theoretical and Mathematical Physics. A.S. would like to thank   the organizers of the XXVIth  International Conference on Integrable Systems and Quantum symmetries (ISQS-26) for their hospitality.

\end{document}